\documentclass[prl,twocolumn,showpacs]{revtex4}
\usepackage[dvips]{graphicx} 
\begin{document}
\title{Number and length of attractors in a critical Kauffman model with
connectivity one}
\author{Barbara~Drossel, Tamara~Mihaljev, Florian~Greil} 
\affiliation{Institut f\"ur Festk\"orperphysik,  TU Darmstadt,
Hochschulstra\ss e 6, 64289 Darmstadt, Germany }
\date{\today}
\begin{abstract}
The Kauffman model describes a system of randomly connected nodes with
dynamics based on Boolean update functions. Though it is a simple
model it exhibits very complex behavior for ``critical'' parameter
values at the boundary between a frozen and a disordered phase, and is
therefore used for studies of real network problems.  We prove here
that the mean number and mean length of attractors in critical random
Boolean networks with connectivity one both increase faster than any
power law with network size. We derive these results by generating the
networks through a growth process and by calculating lower bounds.
\end{abstract} 
\maketitle

Boolean networks are often used as generic models for the dynamics of
complex systems of interacting entities, such as social and economic
networks, neural networks, and gene or protein interaction networks
\cite{kauffman:random}. The simplest and most widely studied of these
models was introduced in 1969 by Kauffman \cite{kauffman:metabolic} as
a model for gene regulation.  The system consists of $N$ nodes, each
of which receives input from $K$ randomly chosen other nodes. The
network is updated synchronously, the state of a node at time step $t$
being a Boolean function of the states of the $K$ input nodes at the
previous time step, $t-1$.  The Boolean updating functions are
randomly assigned to every node in the network, and together with the
connectivity pattern they define the realization of the network. For
any initial condition, the network eventually settles on a periodic
attractor. Thus the number and the lengths of the attractors are
important features of the networks. Of special interest are
\emph{critical} networks, which lie at the boundary between a frozen
phase and a chaotic phase \cite{derrida:random,derrida:phase}.  In the
frozen phase, a perturbation at one node propagates during one time
step on an average to less than one node, and the attractor lengths
remain finite in the limit $N\to \infty$. In the chaotic phase, the
difference between two almost identical states increases exponentially
fast, because a perturbation propagates on an average to more than one
node during one time step \cite{aldana-gonzalez:boolean}. Based on
computer simulations, the mean attractor number of critical $K=2$
Kauffman networks with a constant probability distribution for the 16
possible updating functions was once believed to scale as $\sqrt{N}$
\cite{kauffman:metabolic}. With increasing computer power, a faster
increase was seen (linear in \cite{bilke:stability}, ``faster than
linear'' in \cite{socolar:scaling}, stretched exponential in
\cite{bastolla:relevant,bastolla:modular}). Finally, in a beautiful
analytical study, Samuelsson and Troein
\cite{samuelsson:superpolynomial} have proven that the number of
attractors grows indeed faster than any power law with the network
size $N$.  Concerning the scaling behavior of the mean attractor
length, there is not yet a conclusive result in the literature. While
it appeared to increase as $\sqrt{N}$ in earlier times
\cite{kauffman:metabolic,bhattacharjya:power-law}, Bastolla and Parisi
suggest that it might in fact increase faster than any power law
\cite{bastolla:relevant,bastolla:modular}, and a recent review article
treats this as an open question \cite{aldana-gonzalez:boolean}. Just
as for the attractor number, computer simulations are hampered by
undersampling, which makes it virtually impossible to find attractors
that occur only in few realizations or that have a small basin of
attraction.

In this letter, we want to prove that for $K=1$ critical networks the
mean number of attractors as well as their mean length grows faster
than any power law with the network size.  In such a network out of
the four possible updating functions only the two non-constant ones
occur.  These networks are critical because a perturbation at one node
propagates during one time step on an average to one node.  This is
the first analytical demonstration that in the same ensemble of
networks both these quantities increase faster than any power law. If,
as widely believed, all critical Boolean networks behave in a similar
way, these results should also hold for the critical $K=2$
networks. They can in fact directly be applied to the $\mathcal{B}_1$
class of the $K=2$~critical networks, that is to the $K=2$ ~networks
with only those four Boolean functions that are canalizing and depend
only on one of the two inputs. These networks are equivalent to
critical $K=1$ networks since for $\mathcal{B}_1$ networks the
effective number of inputs per node is 1, as the other input does not
influence the network dynamics.

The topology of networks with $K=1$ consists of loops and of trees
rooted in them, and several exact results for these networks have been
obtained by Flyvbjerg and Kj\ae{}r \cite{flyvbjerg:exact}. The
dynamics on the loops determines the dynamics on the entire network,
and the dynamics on the trees is slaved to the dynamics on the loops.
Relevant nodes are those nodes that are not frozen and that control at
least one other relevant element \cite{bastolla:modular}.  If all 4
update functions are chosen with a nonzero probability, only short
loops have a non-vanishing probability of not containing a constant
function \cite{flyvbjerg:exact}.  Thus the number of relevant elements
remains finite in the limit of infinite network size, and these
networks are always in the frozen phase \cite{bastolla:modular}.
Choosing only non-constant Boolean functions in $K=1$ networks makes
all nodes on all loops relevant, and all possible states on a loop are
part of a cycle in state space.  There are two kinds of non-constant
Boolean functions of one Boolean element: tautology ($\oplus$
coupling) and contradiction ($\ominus$-coupling). The number and the
lengths of the cycles on a loop depend only on the parity of the
number of $\ominus$-functions and not on the details of the
distribution of the Boolean functions. The cycle lengths of an
``even'' loop of length $l$ are 1, $l$, and divisors of $l$. The
maximum cycle length of an ``odd'' loop is $2l$
\cite{flyvbjerg:exact}.

Let us first show that the mean attractor number increases faster than
any power law with~$N$. Let $n_l$ be the number of loops of length
$l$, and $m = \sum_{l=1}^{N}n_l \, l$ the number of nodes in the
loops. $m$ is related to the attractors via
\begin{eqnarray}
\sum_{i} \nu_i A_i &=& 2^{m} \, , \label{sumnA}
\end{eqnarray}
where $\nu_i$ is the number of  attractors of  length $A_i$.  From
here we see that finding an upper bound for the attractor length gives
us a lower bound for the attractor number.

The attractor length~$A$ is the least common multiple of cycle lengths
(periods) of the loops,
\begin{eqnarray}
A \leq A_{\rm max} &=& {\rm LCM} (2l_1, 2l_2, \ldots) \leq 2 \prod_{i}l_i \, .
\end{eqnarray}
For a fixed $m$, this product reaches its maximum if all $l_i$~are equal,
$l_i=l \, \forall i $. 
In this case we have $m = n_l\,l$ and $A < 2 \, l^{n_l}$. 
Maximizing this product as a function of the number~$n_l$ of loops of
length~$l$
\begin{eqnarray*}
\frac{\rm d}{{\rm d} n_l} \left( 2\,l^{n_l} \right) &=& 2 \, 
\frac{\rm d}{{\rm d} n_l} \left( \frac{m}{n_l} \right)^{n_l} =0
\end{eqnarray*}
we obtain
\begin{eqnarray}
A_{\rm max} \leq 2\prod_{i} l_i &\leq& 
2\exp \left( \frac{m}{e} \right) =2 \cdot 2^{0.53m} . \label{upperA}
\end{eqnarray} 
A slightly better upper bound of the form~$2^{0.5m}$ was derived
in \cite{flyvbjerg:exact}, using a much more complicated calculation. From
Eqs.~(\ref{upperA}) and (\ref{sumnA}), we obtain a lower bound for the
number~$\nu_i$ of attractors,
\begin{eqnarray}
\sum_i \nu_i &\geq& \frac{1}{2}\cdot 2^{0.47m}\, .
\end{eqnarray}
Averaging over the different network realizations gives
\begin{eqnarray}
\overline{\sum_i \nu_i} &\geq& \frac{1}{2} \cdot \overline{2^{0.47m}}
\geq \frac{1}{2} 2^{0.47\overline{m}}\, . \label{meann}
\end{eqnarray}
An analytical expression for~$\overline{m}$ can be derived
from the exact results in \cite{flyvbjerg:exact}. 
One of them is the expectation value for the mean number 
of the loops of the length~$l$ in the large $N$ limit
\begin{eqnarray}
\overline{n_l} &=& \frac{1}{l}\exp \left( \frac{-l^2}{2N} \right) \, .
\end{eqnarray}
This result follows also from our Eqs.\,(\ref{LimitingCase})
and (\ref{Nt}) below.  Thus we find for the mean value of the number
of nodes that are on loops
\begin{eqnarray*}
\overline{m} &=& \sum_{l=1}^N \overline{n_l} \,l = 
\sum_{l=1}^N\exp \left( \frac{-l^2}{2N}\right) \, .
\end{eqnarray*}
Approximating this sum with an integral one obtains
$\overline{m} \approx \sqrt{\frac{\pi}{2} \, N}$. 
Inserting this expression in Eq.~(\ref{meann}), we see that the mean number of attractors grows at
least as fast as $\exp \left( 0.4\sqrt{N} \right)$ with the number of nodes.

Next, we show that the mean attractor length diverges faster than any
power law. For this purpose, we generate the ensemble of all
realizations of networks of size $N+1$ via a growth process from the
ensemble of networks of size $N$. The following rule ensures that each
network of the new ensemble is generated exactly once.
The nodes are distinguishable and numbered in the sequence in which they
were added. 
To every network of the initial ensemble we insert a 
new node and add a new link. The new node has either itself as input 
or is linked to a node from the already existing network. 
Next, we have to assign to this new node all
possible combinations of outgoing links. 
This is done such that all possible combinations of the predecessor's
outgoing links can become the outgoing links of the new node. 
If the new node is connected to itself any combination of outgoing 
links of node number~1 that are not on a loop can be moved to it.  
Different rearrangements of these links are
weighed equally and every such network belongs to the new ensemble of
the networks with $N+1$ nodes. This procedure guarantees that the
number of inputs per node of the already existing network is not
changed.  If the node being the one linked to the inserted node was on
the loop, there is a probability of $\frac{1}{2}$ that the new node is
going to be on the loop (since $\frac{1}{2}$ is the probability that
the predecessor's outgoing link that was the part of the loop is
shifted to the new node). 
One can see that the already
existing loops become bigger with time, and that new loops with only
one node are created. We now consider the growth of the networks as
a dynamical process, and we focus only on the loops. Since every node
has the same distribution of numbers and sizes of trees connected to
it, all nodes in all loops become connected to a new node with the
same probability in the ensemble. We define the time scale such that the
rate of insertion of new nodes at a given position in a given loop is
unity,
\begin{eqnarray}
{\rm d}t = \frac{1}{2N}{\rm d}N &\Rightarrow& t=\ln \sqrt{N}\, .\label{Nt}
\end{eqnarray}
Note that $N$ now denotes the mean network size in the growing
ensemble. By going from exact insertion numbers to insertion rates, we
have made a transition to a ``grand canonical'' ensemble. Within this
ensemble, a loop of size $l$ becomes a loop of size $l+1$ with
probability $ldt$ during a time interval $dt$. We then obtain the
following equations for the mean number~$\overline{n_l}$ of loops of
size~$l$
\begin {eqnarray*}
\frac{\rm d}{{\rm d}t} \overline{n_1} &=& 1- \overline{n_1} \\
\frac{\rm d}{{\rm d}t} \overline{n_l} &=&
(l-1) \overline{n_{l-1}} - l \, \overline{n_l} \, \forall \, l> 1 \\
\end {eqnarray*}
With the initial condition $N(0)=1$, these equations have the solution
\begin{eqnarray*}
\overline{n_1}(t)&=& 1 \\
\overline{n_2}(t)&=&\frac{1}{2} - \frac{1}{2}e^{-2t} \\
\overline{n_3}(t)&=&\frac{1}{3} - e^{-2t} + \frac{2}{3}e^{-3t} \\
\overline{n_4}(t)&=&\frac{1}{4} - \frac{3}{2} e^{-2t} + 2 e^{-3t} - \frac{3}{4}e^{-4t} \\
&\vdots&
\end{eqnarray*}
For the limiting case of large times, this solution can be approximated by
\begin{eqnarray}
\overline{n_l}(t) &=& \frac{1}{l}- \frac{l-1}{2} e^{-2t} 
\label{LimitingCase}
\end{eqnarray}
which approaches the stationary solution $\overline{n_l}=l^{-1}$ for ${\mbox t \to \infty}$.
Introducing the small parameter $\epsilon \ll 1$ as a measure of how far we are from the stationary solution, we find from
\begin{eqnarray*}
\overline{n_{l_c}} (t)= \frac{1}{l_c}- \frac{l_c-1}{2} e^{-2 \ln\sqrt{N}} &=&
\left( 1-\epsilon \right) \frac{1}{l_c}
\end{eqnarray*}
that the critical value for the loop size for large~$\epsilon N$ is 
\begin{eqnarray}
l_c &=& \sqrt{2 \epsilon N}\, .
\end{eqnarray}
In the same manner we may write the master equation for the probability 
distribution~$P(n_1, \ldots, n_l)$ of the loops smaller than $l$,
\begin{eqnarray*}
\frac{\rm d}{{\rm d}t}P(n_1,\dots,n_l) = 
- \sum_{i=1}^li\;n_i\;P(n_1,\dots,n_i,n_{i+1},\dots,n_l) &&\\
 + \sum_{i=1}^l (i-1)(n_{i-1}+1)\;P(n_1,\dots,n_{i-1}+1,n_{i+1},\dots,n_l)
\end{eqnarray*}
The stationary solution for this expression valid for the loops smaller than $l_c$ is
\begin{eqnarray}
P(n_1,\dots,n_l) &=& \prod_{i=1}^l e^{-1/i} 
\left( \frac{1}{i}\right)^{n_i}\frac{1}{n_i!} \, .
\label{statsol}
\end{eqnarray}
This solution is time independent and we can conclude that the
distribution of the loops smaller than $l_c$ is not changing with
time, i.e.\,with the growth of the system size. Furthermore the
probabilities for having $n_i$ loops of size $i$ are independent from
each other and Poisson distributed with a mean $i^{-1}$. 

Equipped with these results, we can now evaluate the lower bound for
the mean attractor length $\overline{A}$. Suppose that the system is
enlarged so that its number of nodes is $N'=aN$, with $a>1$ and
$N$~nodes of the previous system. The length of the attractor is the
least common multiple of the loop periods, i.e.\,the cycle lengths of
the loops. Since $\overline{A_{l_i\leq N'}} \geq \overline{A_{l_i\leq
l_c(N')}}$, we obtain a lower bound by evaluating only the change of
the least common multiple of the periods of loops smaller than $l_c$,
that is the change of $\overline{A_{l_i\leq l_c}} \equiv
\overline{A_{\sqrt{2 \epsilon N'}}}$, with increasing system size. Our
above considerations show that the distribution of loops of size
smaller than $l_c(N)$ does not change when going to an ensemble of
systems of size $N'$. However, these systems contain additional loops
in the interval $[l_c(N),l_c(N') ]$.  If the period of such an
additional loop is a prime number larger than $l_c(N)$, the least
common multiple of all loop periods is multiplied by this period.  A
loop with a prime number of nodes has only two possible periods: 1 or
$l$ if the loop is even, and 2 or $2l$ if the loop is odd. If the
additional loop is not on the cycle of length 1 or 2, the least common
multiple of the periods of the loops smaller than $l_c(N')$ is at
least as large as the product of the new loop size, and the least
common multiple of the periods of the loops smaller than $l_c(N)$.
The number of primes not exceeding the value of some positive number
$x$ is asymptotically expressed as $\pi_x=x/ \ln x $ (see,
e.g.\,\cite{hardy:introduction}).  The probability that a randomly
chosen number in the interval $[l_c(N),l_c(N') ]$ is a prime number is
\begin{eqnarray}
P_{\rm prime} &=& 
\frac{\pi_{\sqrt{2\epsilon N'}}-\pi_{\sqrt{2\epsilon N}}}{\sqrt{2\epsilon N'}-\sqrt{2\epsilon N}}
\approx \frac{1}{\ln\;\sqrt{2 \epsilon N'}} \, .
\end{eqnarray} 
This is identical to the probability that the new loop size is a prime number. 
Taking all these considerations together, we have
\begin{eqnarray*}
\overline{A_{l_i\leq N'}} \geq \overline{A_{\sqrt{2 \epsilon N'}}}
&\geq&  P_{\rm loop}P_{\rm not 1,2}\frac{1}{\sqrt{a}} 
\frac{\sqrt{2 \epsilon N'}}{\ln (\sqrt{2 \epsilon N'})} \overline{A_{\sqrt{2 \epsilon N}}}\,.
\end{eqnarray*}

The probability $P_{\rm loop}$ for having a loop in the interval $[l_c(N),l_c(N') ]$ is obtained using (\ref{statsol}). 
The probability of having no loop of size $l$, $n_l=0$, is $e^{-1/l}$. 
Thus the upper bound for the probability of having no loops with size from 
the interval $[l_c(N),l_c(N') ]$  is
\begin{eqnarray*}
1-P_{\rm loop} &\leq& \prod_{i=l}^{a\, l} e^{-\frac{1}{i}} 
\leq\left( \exp\left( -\frac{1}{al}\right) \right) ^{al-l+1}\\
&=& \exp\left( \frac{1-a}{a}-\frac{1}{al}\right) \stackrel{0\leq l^{-1}\leq 1} \leq e^{-1}
\end{eqnarray*}
giving
\begin{eqnarray}
P_{\rm loop}\geq 1-e^{-1}\, . 
\end{eqnarray}
The probability $P_{\rm not 1,2}$ that the new loop is not on an
attractor of length 1 or 2 is obtained as follows: The number of its
cycles is $(2^l-2)/l+2$ in the case of the even loops  
and $(2^l-2)/2l+1$ for the odd loops. 
Among these cycles two are of length 1 for the first type of loop and one is of length~2 for 
the second type. 
The probability that the loop of size $l$ is not on a cycle of
length 1 or 2 is $1-2l/2^l$ for large values of $l$. 
The loop we are observing is of size $\sqrt{2 \epsilon N}$ and for 
the probability~$P_{\rm not 1,2}$ we obtain
\begin{eqnarray}
P_{\rm not 1,2} &=& 1-\frac{2\sqrt{2 \epsilon N}}{2^{\sqrt{2 \epsilon N}}}.
\end{eqnarray} 
For a given $\epsilon$ this probability is non-vanishing, 
i.e.\,$P_{\rm not 1,2}>\eta>0$, if $N>2/\epsilon$. 
Since we are considering the limit of large $N$, this condition is satisfied.  
Applying this result to the lower bound for the attractor size we finally have
\begin{eqnarray}
\overline{A_{\sqrt{2 \epsilon N'}}} &\geq&
(1-e^{-1})\eta \frac{1}{\sqrt{a}} \frac{\sqrt{2 \epsilon N'}}{\ln\; \sqrt{2 \epsilon N'}}
\overline{A_{\sqrt{2 \epsilon N}}}\, . \label{finres}
\end{eqnarray}
Setting $N=a^\mu N_0$ and defining a constant~$C$
\begin{eqnarray*}
C &=& \left( 1-e^{-1} \right) \frac{\eta \sqrt{2 \epsilon}}{\ln(\sqrt{2 \epsilon aN_0})} 
\left( \frac{N_0}{a} \right)^{1/4} 
\end{eqnarray*}
Eq.~(\ref{finres}) can be transformed into
\begin{eqnarray*}
\overline{A_{l_i\leq N}} &\geq& \frac{C^\mu N^{\mu/4}}{\mu!}\, ,
\end{eqnarray*}
and finally with $\mu= \left(\ln(N/N_0)/\ln (a)\right)$ into
\begin{eqnarray}
\overline{A_{l_i\leq N}} &\geq& 
\left(\frac N N_0\right)^{\frac{4\ln C + \ln N}{4\ln a}}
\!\!\!\!\!\!\!\!\!\!\!\! 
\frac 1{(\ln(N/N_0)/\ln a)!} \overline{A_{\sqrt{2 \epsilon N_0}}}\, .\label{end}
\end{eqnarray}
This increases faster than any power law with $N$, but slower than a
stretched exponential.

It remains to see in how far these results apply also to critical $K=2$
Kauffman networks. Bastolla and Parisi \cite{bastolla:modular} have pointed out
that the set of relevant nodes of these networks consists of modules,
which together determine the attractor numbers and lengths. This
situation is very similar to the critical $K=1$ networks treated in
this paper, where the modules built of relevant nodes are loops, the
properties of which determine the attractors. The main difference in
$K=2$ networks is that the modules are more complicated. A thorough
evaluation of their properties has not yet been done. 

Nevertheless, the evidence cited so far suggests that for all critical
Kauffman networks the mean number and length of attractors diverges
faster than any power law. This means that the attractors are too long
and too many to represent cellular differentiation, to which the model
was originally applied. The vast number of attractors in these models
appears to be a consequence of the synchronous updating scheme. Recent
studies of modified models that allow for randomness in the updating
rules, indicate that a deviation from synchronous update reduces the
number of attractors considerably \cite{klemm:topology}, which now
becomes a power law \cite{klemm:stable, greil:dynamics}. However, in order
to model biological networks realistically, further modifications are
needed, and the present work is only a small step on the long way
towards understanding regulatory networks.

\begin{acknowledgments}
We thank Viktor Kaufman for useful discussions.
\end{acknowledgments}

\bibliography{PRL_k1}
\end{document}